\documentclass[pra,twocolumn]{revtex4}
\usepackage{graphicx}

\begin{document}

\title{Spatial interference from well-separated condensates}

\author{M.~E.\ Zawadzki}
\author{P.~F.\ Griffin}
\author{E.\ Riis}
\author{A.~S.\ Arnold}
    \affiliation{SUPA, Dept.\ of Physics, University of Strathclyde, Glasgow G4 0NG, UK}
    \homepage{www.photonics.phys.strath.ac.uk/AtomOptics/}

\date{\today}

\begin{abstract}
We use magnetic levitation and a variable-separation dual optical plug to obtain clear spatial interference between two condensates axially separated
by up to $0.25\,$mm -- the largest separation observed with this kind of interferometer. Clear planar fringes are observed using standard (i.e.\
non-tomographic) resonant absorption imaging. The effect of a weak inverted parabola potential on fringe separation is observed and agrees well with
theory.
\end{abstract} \maketitle

It is now over a decade since Andrews \textit{et al.}'s \cite{KettScience} impressive demonstration of the wavelike nature of coherent matter via the
spatial interference of $^{23}$Na Bose-Einstein condensates (BECs). Such matter-wave interference experiments are of great interest for applications
in ultra-precise interferometry \cite{refsensing}, and should lead to drastic improvements in measurements of fundamental constants as well as
temporal, gravitational and rotational sensing. Here we obtain spatial BEC interference which promises significant potential for improved
measurements. We use a magnetic levitation field \cite{Grimm} to spatially interfere two atomic clouds with relatively large spatial separations of
$0.25\,$mm. Moreover we use an atomic species, $^{87}$Rb, with 4 times the mass of Ref.~\cite{KettScience} and hence 4 times smaller de Broglie
wavelength for the same atomic velocities. We find tomographic imaging \cite{KettScience} is not required, and standard absorption imaging suffices
for good contrast $60\,\%$ ($30\,\%$) interference at separations of $60\,\mu$m ($250\,\mu$m). We also identify a clear relationship between the
interference fringe period and magnetic levitation time in an inverted parabola trap potential.

\begin{figure}[!t]
    \centering
       \includegraphics[width=.7 \columnwidth]{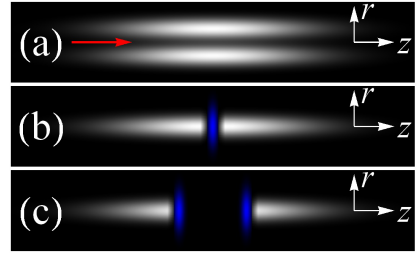}
    \caption{Schematic diagram illustrating different BEC splitting geometries: (a) radial splitting using e.g.\ RF dressed potentials
    (red arrow indicates imaging direction for fringe observation),
    (b) axial splitting using a blue detuned dipole beam, (c) the variable separation axial splitting using dual dipole beams we report on
    in this paper.\label{fig:schematic}}
 \end{figure}

Experiments on atomic interference have developed rapidly in the last decade and it is now possibly to interfere single particles in quantum walks
using the relative population of atoms in a particular state \cite{qw}. A Ramsey-type BEC interferometer using Bragg scattering has also obtained the
largest time-integrated separation in condensate interference experiments \cite{sackett}. Similar advances have so far been unobtainable with
`Young-type' spatial interference patterns, in which the de Broglie waves of two expanding wavepackets, initially spatially separate, give rise to
the interference. Condensate wavefunction irregularities and vortices are only observable with such spatial interferometers. Recently radial
splitting of condensates \cite{rfdress} using RF dressed potentials \cite{barry} has become popular (Fig.~\ref{fig:schematic} (a)), as high contrast
spatial interference fringes can be obtained due to the `point source'-like properties of the condensates when viewed along the BEC axis. Note,
however, that in the radial splitting geometry typical chip BECs can only yield interference patterns for split distances up to $26\,\mu$m
\cite{kettsepar}, or times around $400\,$ms (for $9\,\mu$m separation) \cite{ketttime}. Here we split our cigar-shaped BEC with a far-detuned optical
dipole laser beam which propagates perpendicular to the BEC's longitudinal axis (Fig.~\ref{fig:schematic} (b)), a geometry similar to
Ref.~\cite{KettScience}, where interference from $40\,\mu$m BEC separation was obtained. We use a dual optical plug (Fig.~\ref{fig:schematic} (c)) to
extend our condensate separation from $60\,\mu$m to $250\,\mu$m and back, observing a visibility of 30\% after an experimental time of $300\,$ms.

It should be stressed that if two independent condensates are formed (as in this experiment), or the splitting period is too long relative to the
difference in chemical potentials of the two condensates, then the interference pattern has a random phase \cite{KettScience,kettsepar}. For
practical interferometric applications a single condensate must be smoothly split into two condensates with a fixed relative phase
\cite{rfdress,ketttime,shin}. In future we intend to extend our proof-of-principle interferometry into the phase-coherent regime.

The Bose-Einstein condensate was created in the experimental setup described in detail in Ref.~\cite{ring}. Our Ioffe-Pritchard magnetic trap has
frequencies of $10\,$Hz and $108\,$Hz in the axial and radial direction respectively. Atoms are trapped in the $|F=2,m_F=2\rangle$ trapping state,
with a $40\,$s magnetic trap lifetime and $5\times 10^5$ atoms in a pure BEC. The condensate creation and manipulation was observed by standard
absorption imaging. The imaging beam propagated perpendicular to the BEC axis and a $2\times$ beam expander was used for imaging onto an Andor Luca
CCD camera. The size of individual pixels is $10\,\mu$m, corresponding to $5\,\mu$m at the BEC's location.

Our dipole beam is generated by $50\,$mW of light from a free-running $658\,$nm diode laser, far to the blue of the Rb D2 resonance at
$780\,$nm. In order to create a high intensity dipole beam we used an $80\,$mm focal length achromat lens. The elliptical diode laser beam shape
was focused to beam waists of $8.8\,\mu$m and $13.7\,\mu$m in the axial and radial direction, respectively. This yields a maximum potential of
$30\,\mu$K which completely isolates split condensates from each other and tunneling effects can be neglected. The dipole beam allows fully
coherent and adiabatic splitting of the BEC, with an estimated condensate photon scattering rate of $1\,$mHz per atom. The alignment of the dipole
potential was facilitated by combining the $658\,$nm beam with a `tracer' $780\,$nm repump ($F=1\rightarrow F'=2$) beam \cite{tracer} on a
beamsplitter to create a co-propagating beam with a much higher scattering rate and optical potential.
For all experiments in this paper the dipole beam was on throughout evaporative cooling to BEC -- resulting in the creation of two independent
samples of coherent matter with random relative phase.

As acousto-optical modulators (AOMs) can vary the deflection angle and beam intensity of a dipole beam via the applied RF frequency and power,
respectively, they are a useful tool for creating arbitrary patterns in BEC experiments through the time-averaged optical dipole
potential~\cite{kettsurf,schnelle,malcolm}. However, their use with BECs has largely been through red-detuned light, although blue-detuned potentials
\cite{aidan} offer substantially lower decoherence rates.

We split our blue-detuned dipole laser beam into two beams, with variable separation, via an $80\,$MHz AOM. As the first order beam from an AOM is
deflected proportionally to the RF drive frequency, if we use an RF spectrum consisting of multiple spectral components we can form multiple
simultaneous beams \cite{shin}. Our adiabatic splitting is induced by dipole beam sidebands driven by amplitude modulation of the RF carrier
frequency fed to the AOM. The amplitude modulation is obtained by mixing two frequencies, a stable carrier frequency $\nu_0=80\,$MHz and a variable
frequency modulation signal $0<\nu_{\rm mod}<20\,$MHz, yielding two tunable sidebands at $\nu=\nu_0 \pm \nu_{\rm mod}$. The RF modulation frequency
came from a computer-controlled synthesized signal generator. A standard double balanced mixer is used to mix the signal and carrier RF signals.
The decrease in amplitude of the carrier frequency from the sidebands is of order $40\,$dB, and the carrier frequency dipole beam has a negligible
effect on the atoms. The linear response of RF drive frequency to beam deflection results in two beams at relative deflection angles
$\delta\theta=\pm 2.5\,$mrad for $20\,$MHz modulation frequency.

For small modulation frequencies, the two dipole beams have a good spatial overlap, effectively resulting in a single beam with a beat phenomenon at
the modulation frequency -- i.e. the beam intensity varies bright/dark sinusoidally in time with period $T=1/(2\nu_{\rm mod})$. To highlight the low
heating rate of blue detuned light we used this beating to perform an experiment similar to that of Ref.~\cite{schnelle} -- we studied heating as a
function of intensity modulation frequency of the dipole beam during evaporation to BEC. Heating was observed as the fraction of BEC lost after RF
evaporation in a double-well potential composed of the magnetic trap with a dipole beam that had a sinusoidally modulated intensity. The main result
was that no heating was observed for modulation rates greater than $1\,$kHz, a limit significantly lower than the $30-40\,$kHz of
Ref.~\cite{schnelle}. In principle the trap might be adiabatically deformable at modulation frequencies less than $1\,$kHz, however because of atomic
motion in the harmonic magnetic trap care would then need to be taken that the trap modulation does not interfere with evaporation.



\begin{figure}[!b]
    \centering
        \includegraphics[width=.9\columnwidth]{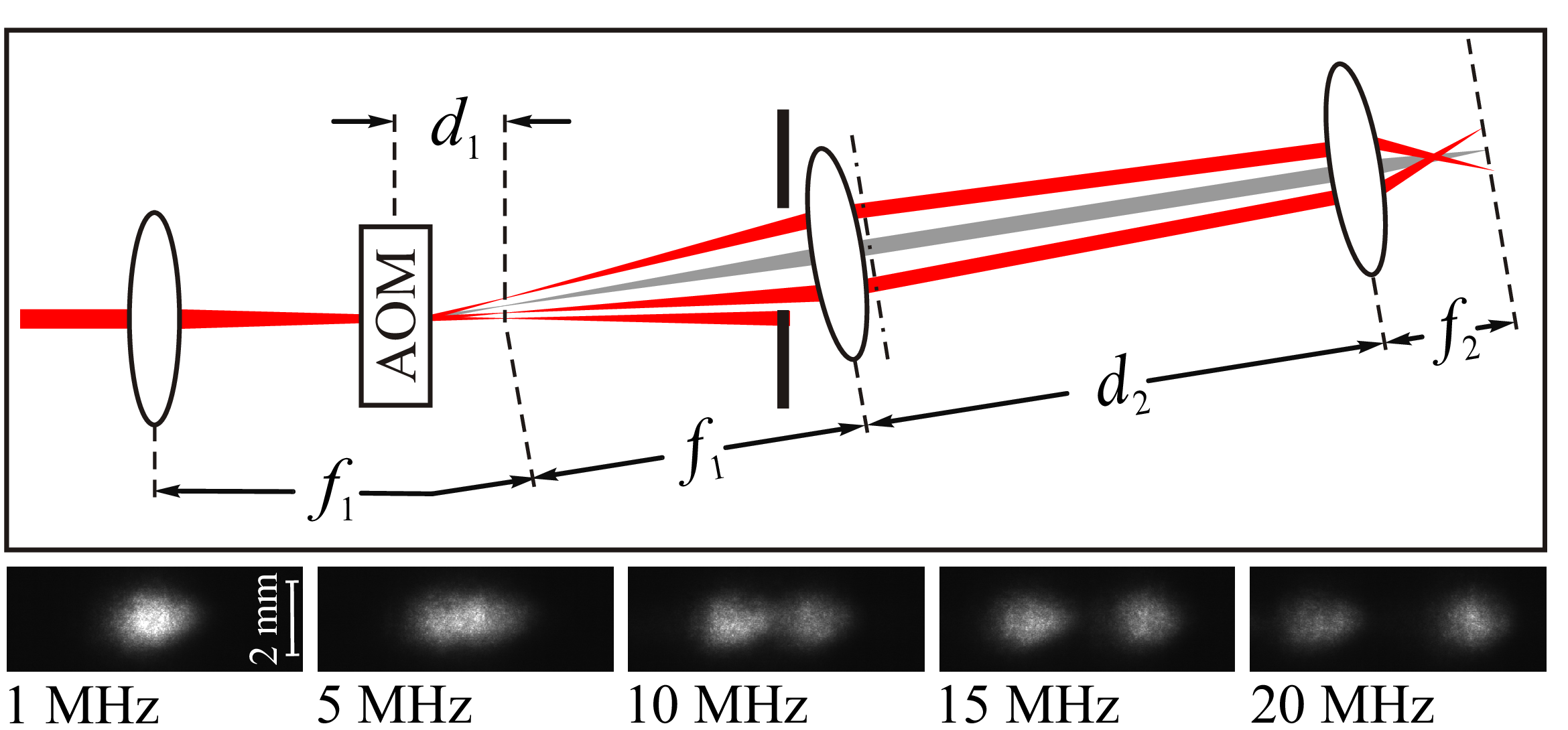}
    \caption{The $658\,$nm dipole beam path for splitting the BEC. The acousto-optical modulator (AOM) was offset a distance $d_1$ from the
    focal point of the 1$\times$ beam expander to enable output beam deflection at the beam waist after the final lens. The modulated RF
carrier frequency results in two RF sidebands and suppressed carrier resulting in two optical beams with spatial separation determined by the RF
modulation frequency as seen in the experimental beam image series.}
    \label{fig:AOM rays}
\end{figure}

The position of our AOM (Fig.~\ref{fig:AOM rays}) was offset by a distance $d_1=10\,$cm from the focal point of a $1\times$ beam expander comprised
of two $f_1=25\,$cm focal length plano-convex lenses. After a (non-critical) propagation distance $d_2$ the beams are focused by an achromat lens
with focal length $f_2=8\,$cm. Using standard paraxial $ABCD$ matrices one can show that the waist after the $f_2$ lens yields beam displacements
$\delta z= d_1\, f_2\,\delta\theta/f_1 = \pm 80\,\mu$m for a modulation frequency of $20\,$MHz (Fig.~\ref{fig:AOM rays}). Although the RF power in
the sidebands is constant a small drop in the optical power of the beams is observable at large displacements due to reduced AOM diffraction
efficiency. The largest achievable center-of-mass separation of two BECs by the repulsive potential of the dipole beams was $250\,\mu$m, with spatial
interference between separated condensates still clearly observable. We believe this is the largest splitting observed in a `Young-type' spatial BEC
interferometer.

The anisotropic character of a cigar-shaped Ioffe-Pritchard trap leads to two different expansion velocities as the mean-field forces from a
repulsive BEC are proportional to the condensate's density gradient - hence the expansion velocity is much greater in the radial direction than the
axial direction. Moreover, by using a dipole beam to create a macroscopic axial separation of our matter waves we need a concomitantly longer
expansion time for BEC recombination and interference than is required for radially split BECs \cite{rfdress}.

The fringe spacing $\lambda$ arises from the de Broglie waves of two condensates and takes the familiar form:
\begin{equation}
\label{deB} \lambda=h/(m v),
\end{equation}
where $h$ is Planck's constant, $m$ is the atomic mass and $v=d/t$ is the \textit{relative} speed between two point-like condensates as a function of
their center-of-mass separation $d$ and expansion time $t$. The duration of ballistic expansion in freefall, $t,$ is usually limited by the size of
the imaging area and the dimensions of the BEC vacuum cell -- times around $100\,$ms lead to long drops of $49\,$mm and the corresponding condensate
speed of $1\,$m/s leads to blurred images. To eliminate the inconvenience of gravity a `levitation' field can be used \cite{Grimm} whereby a magnetic
field gradient counteracts the gravitational acceleration. The levitation field keeps the atoms in the region of interest for time intervals
($t>80\,$ms) which are long enough to make our interference pattern optically resolvable.

Our levitation field is provided by the existing four circular coils which form the toroidal quadrupole field \cite{ring} of our ring Ioffe-Pritchard
trap. The levitation mechanism uses the weak-field-seeking $|2,2\rangle$ atoms of the BEC which are attracted to the local field minimum. After
creation of a BEC in the magnetic trap by a $25\,$s evaporative cooling cycle, `anti-gravity' conditions are obtained with a vertical gradient of
$15\,$G/cm. An additional vertical constant field is added to the quadrupole magnetic field to reduce lensing \cite{bounce} in the vertical and
imaging directions.

\begin{figure}[!th]
    \centering
        \begin{minipage}{.57\columnwidth}
        \includegraphics[width=.99\columnwidth]{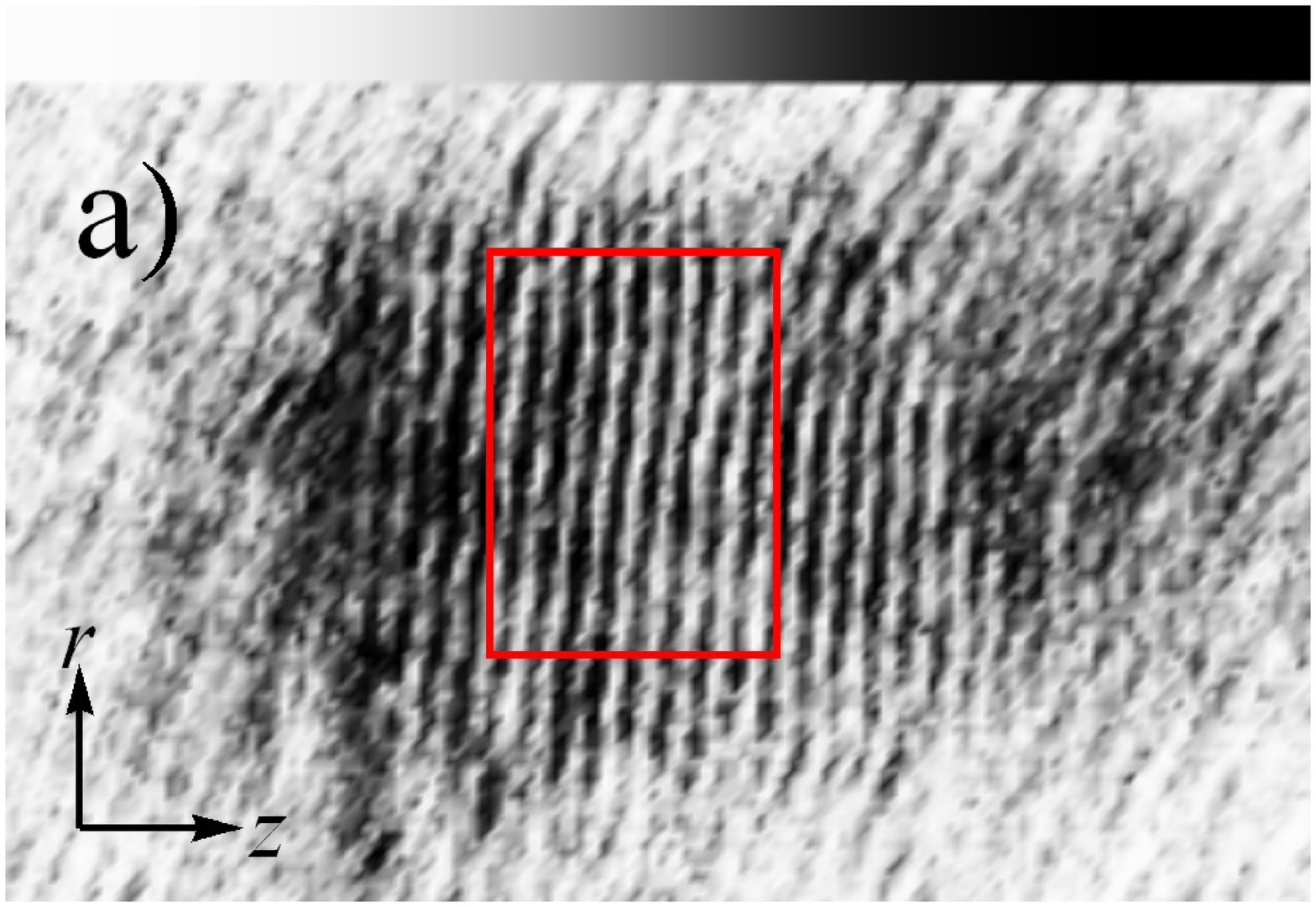}\end{minipage}
                \begin{minipage}{.41\columnwidth}
                            \begin{minipage}{.6\columnwidth}\includegraphics[width=.99\columnwidth]{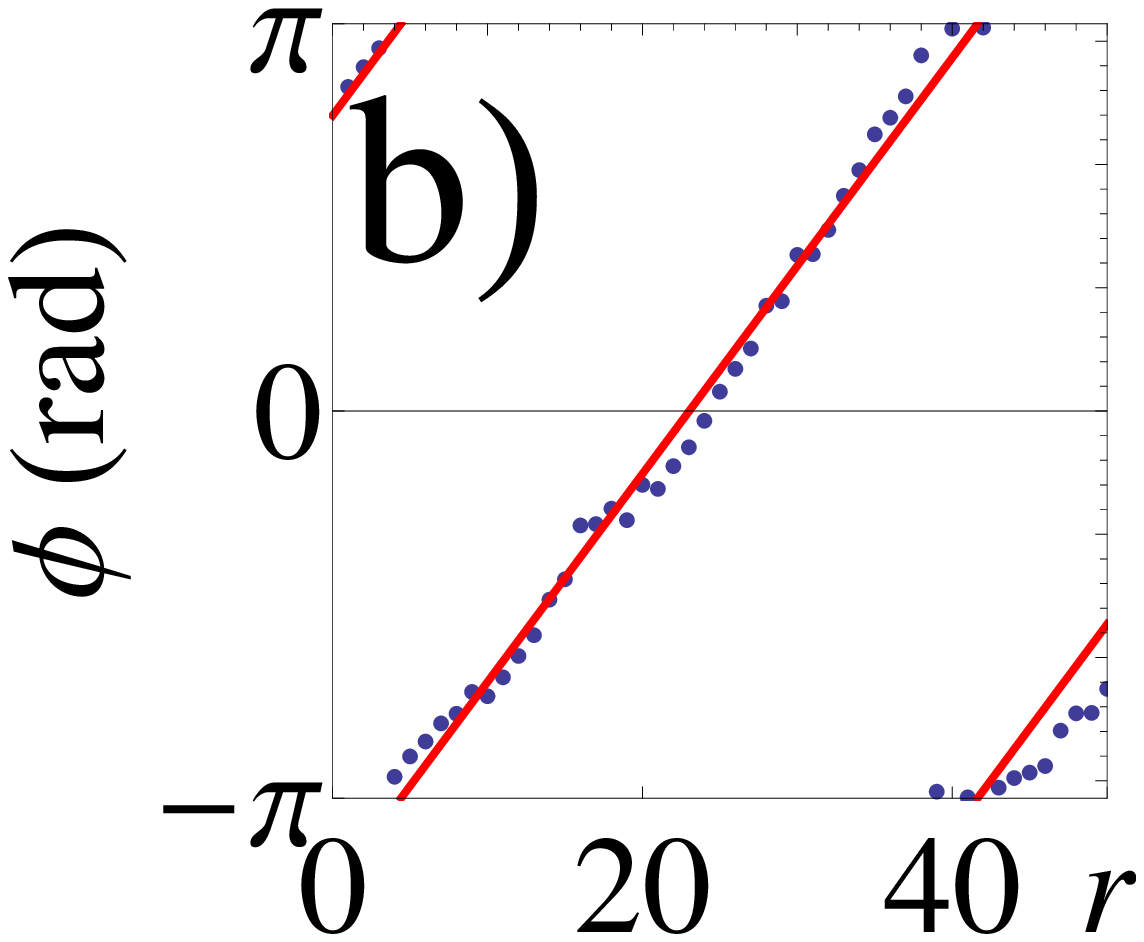}\end{minipage}
            \begin{minipage}{.35\columnwidth}\includegraphics[width=.99\columnwidth]{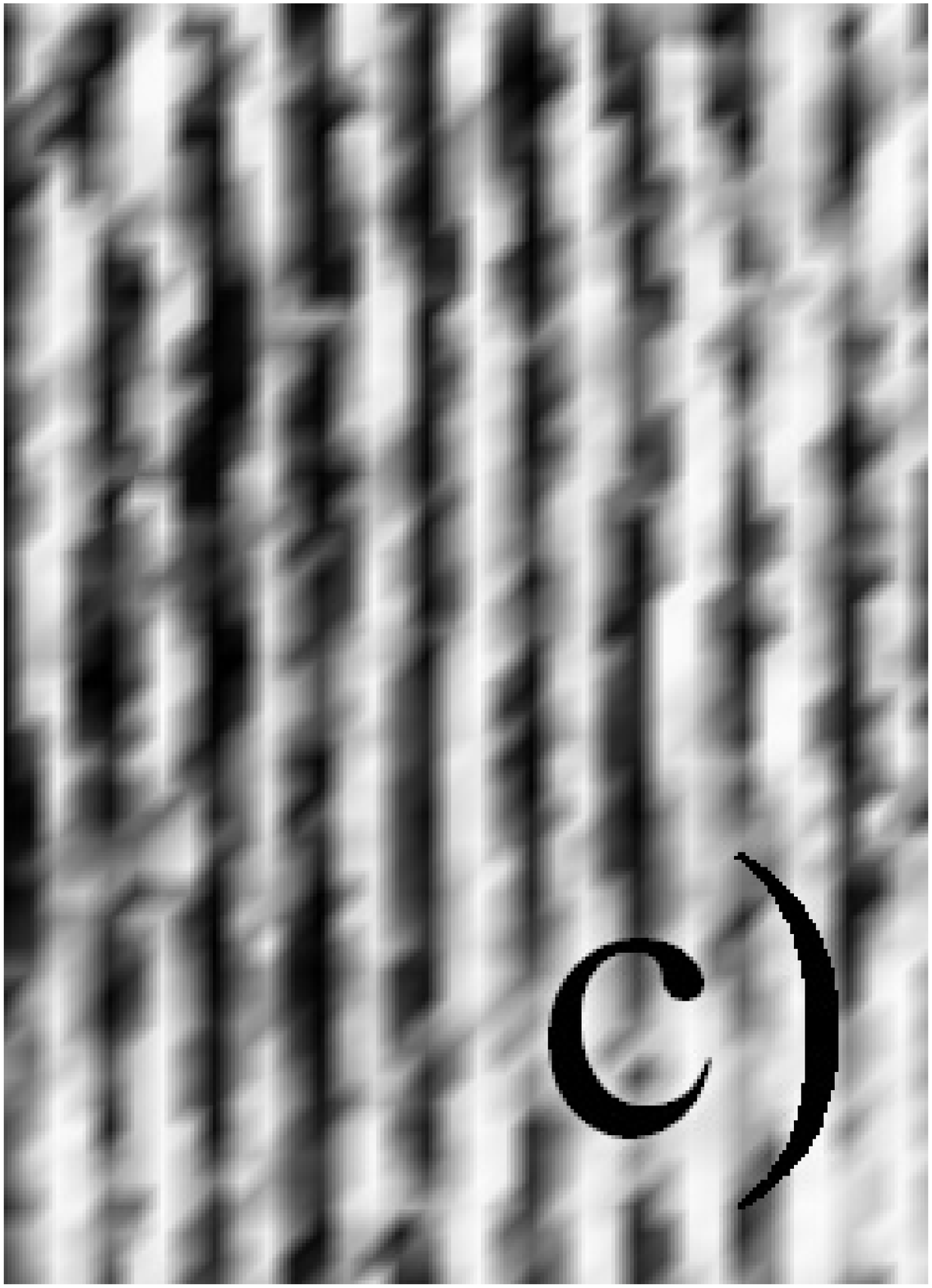}\end{minipage}
        \includegraphics[width=.96\columnwidth]{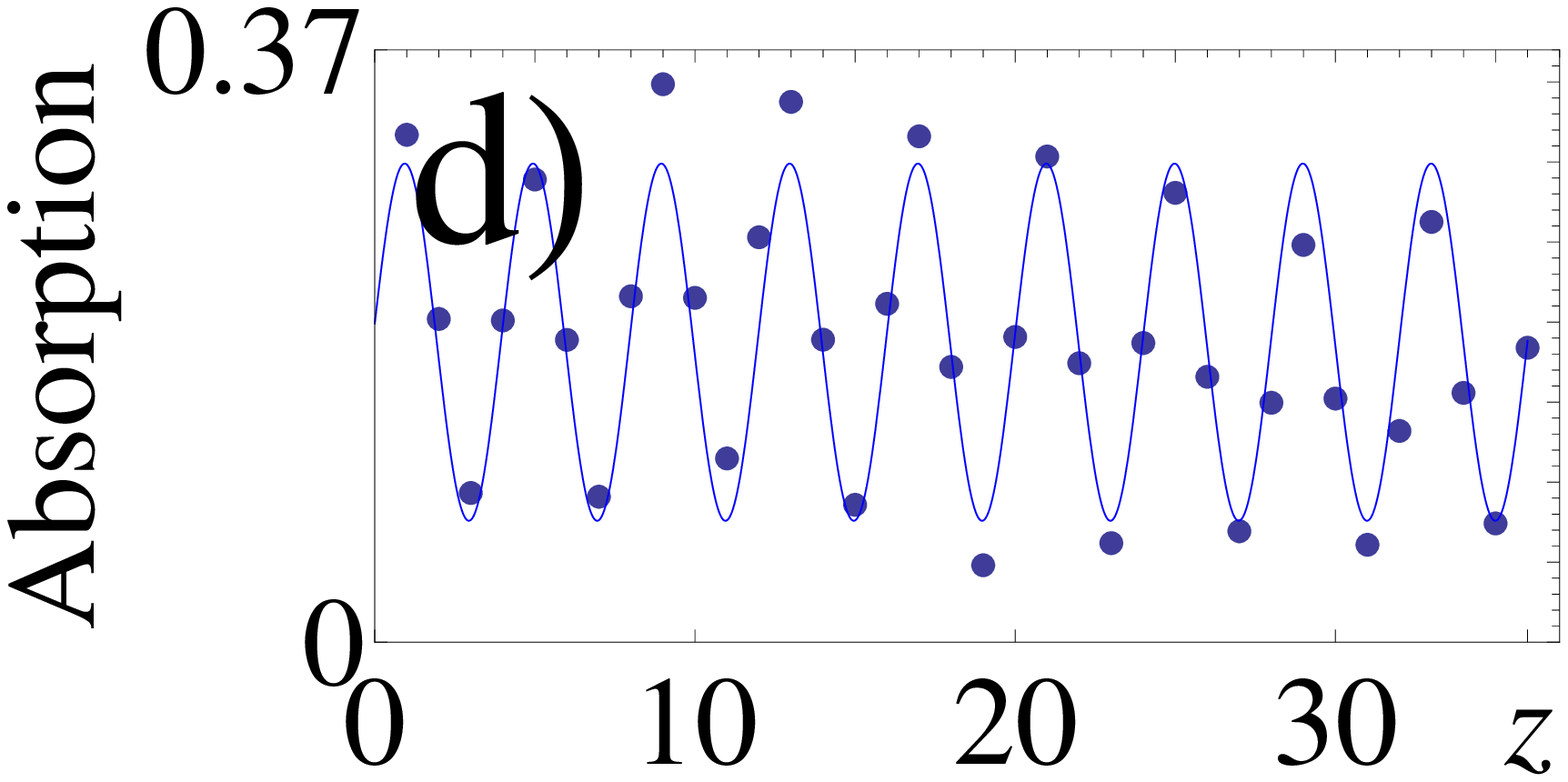}\end{minipage}
        \begin{minipage}{.57\columnwidth}
            \vspace{3mm}    \includegraphics[width=.99\columnwidth]{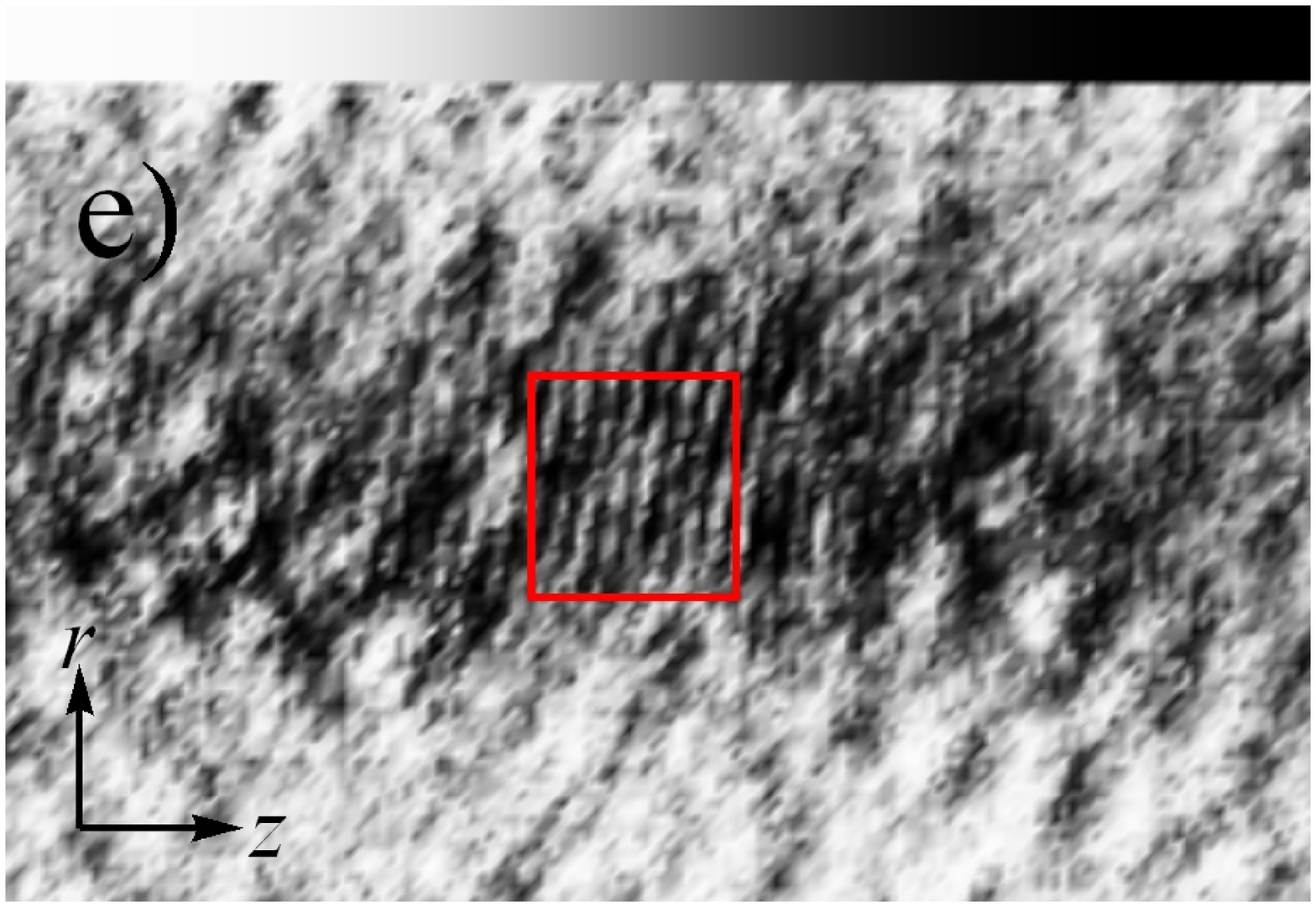}\end{minipage}
                \begin{minipage}{.41\columnwidth}        \vspace{3mm}
                \begin{minipage}{.55\columnwidth}\includegraphics[width=.99\columnwidth]{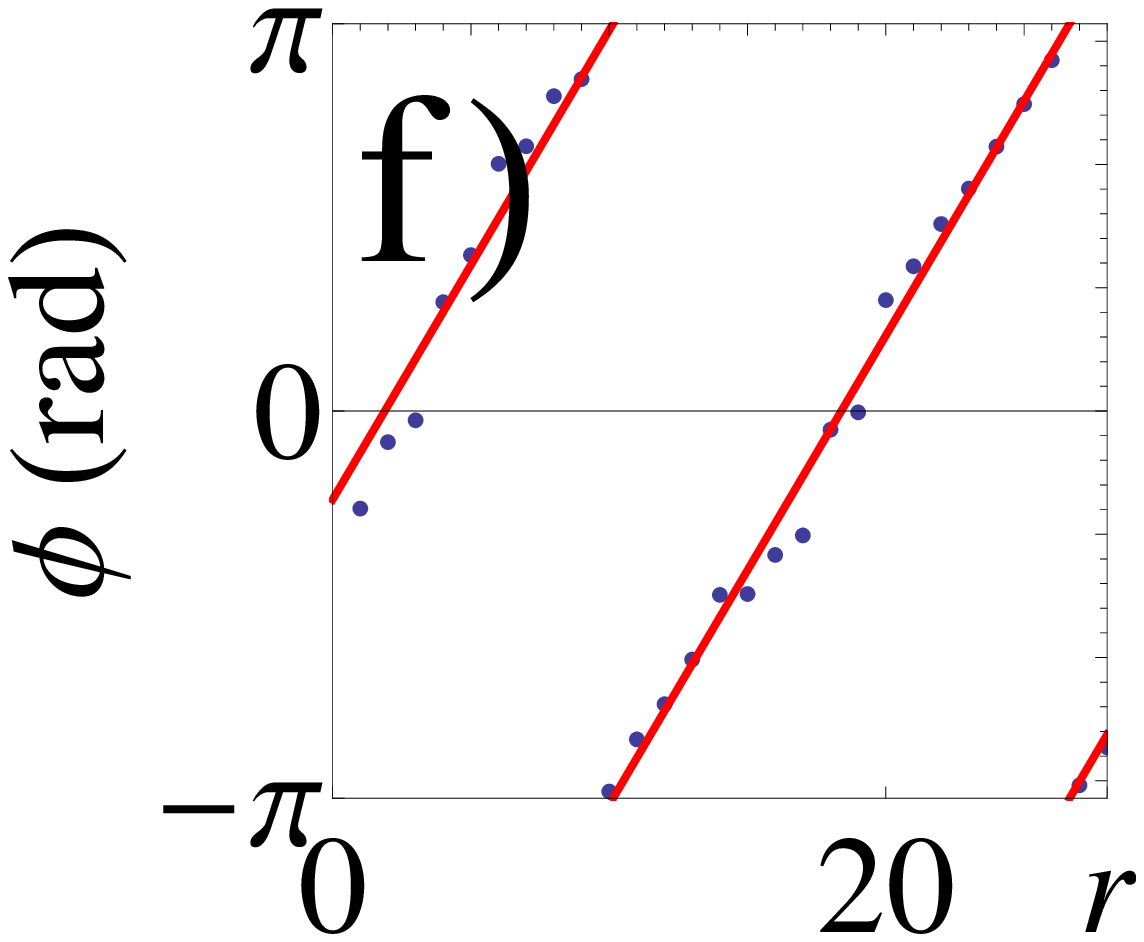}\end{minipage}
            \begin{minipage}{.4\columnwidth}\includegraphics[width=.99\columnwidth]{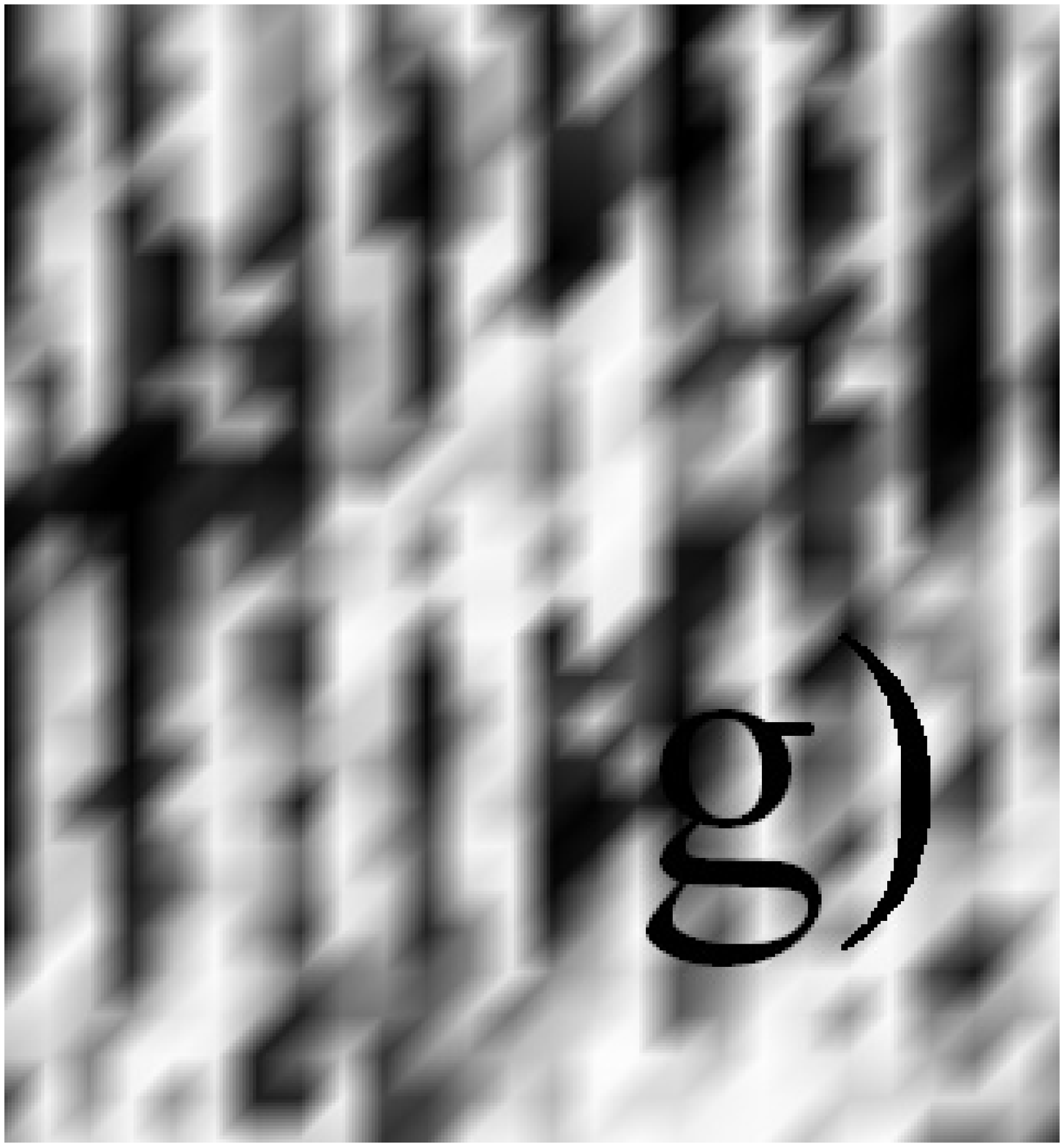}\end{minipage}
        \includegraphics[width=.96\columnwidth]{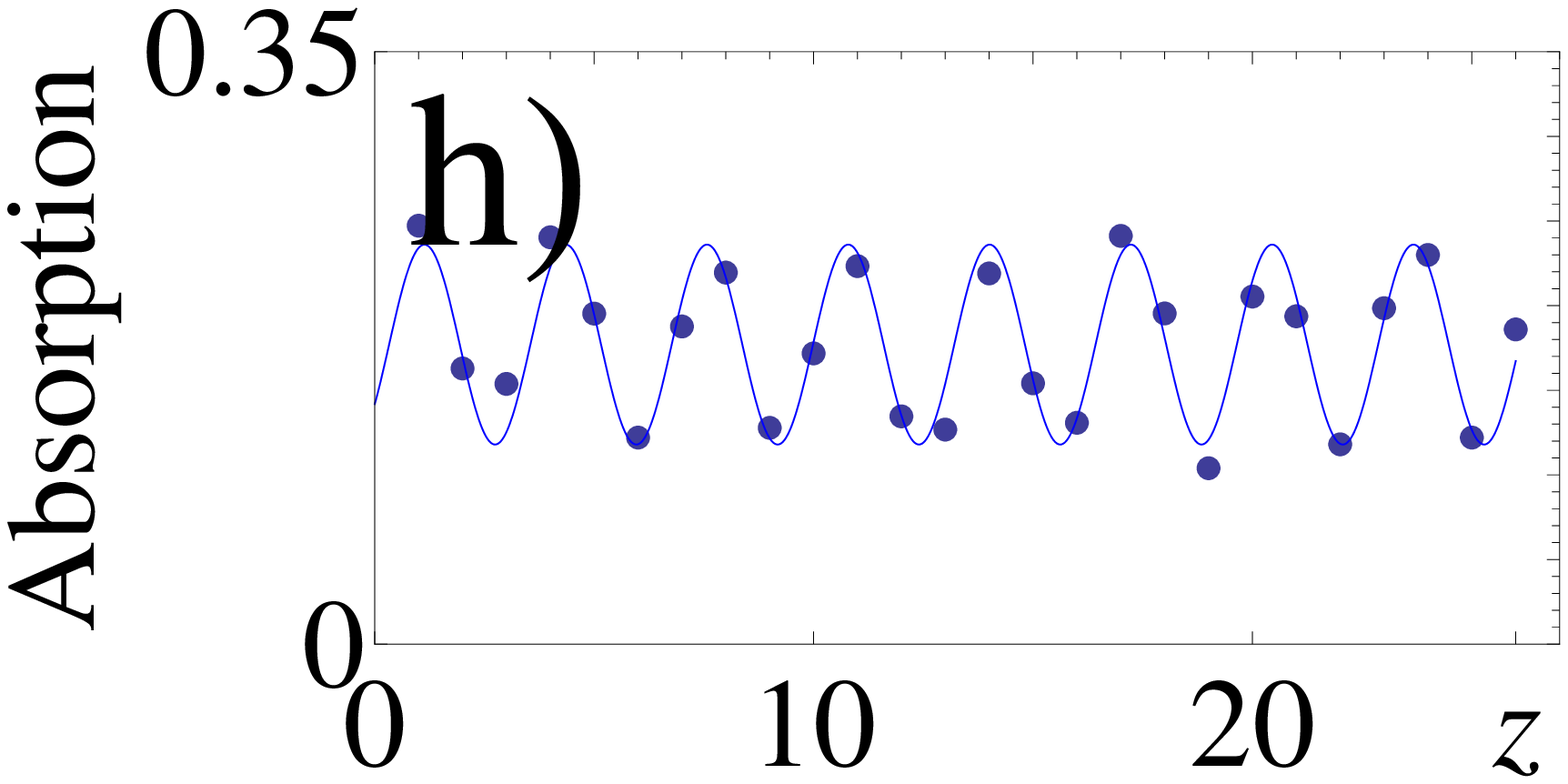}\end{minipage}
    \caption{Interference patterns $(0.8\times0.5\,$mm$^2)$ for: a) $60\,\mu$m separated BECs, e)
    BECs split $60\,\mu$m-$250\,\mu$m-$60\,\mu$m over a $160\,$ms period. In both cases the pictures were taken after a further $135\,$ms of
 magnetic levitation, which for a) corresponds to the triangle point in Fig.~\ref{fig:FringeSpacing}. The phases of the Fourier components of the
 fringes for the red selected areas in a) and e) can be obtained (blue dots in b) and f)), and fit with a sawtooth linear phase shift (red curves).
 These sawtooth phase corrections can then be applied to the Fourier transform, before performing the inverse transform shown in c) and g). These
 corrected images can then be averaged over the image rows to obtain the blue dots in d) and h), with their sinusoidal fits (blue curves). Absorption is
 measured using the natural logarithm. Each row/column (i.e.\ pixel) corresponds to $5\,\mu$m$\times5\,\mu$m. The fringe period in e) is smaller than a) as the condensates have a residual
 counterpropagating velocity after the $250\,\mu$m$\rightarrow\!60\,\mu$m separation phase.  \label{fig:fringes1}}
\end{figure}

Fig.~\ref{fig:fringes1}a) represents an example of the high contrast ($60\%$) interference pattern when two BEC clouds were originally separated by
$60\,\mu$m (center-of-mass (COM) distance) with a single optical plug then recombined using the levitation magnetic field. Standard (i.e.
non-tomographic \cite{KettScience}) absorption imaging is used. The interference pattern when the BEC is split from a COM separation of $60\,\mu$m to
$250\,\mu$m over $80\,$ms, returned to $60\,\mu$m separation over $80\,$ms and then levitated for $150\,$ms has clear continuous spatial fringes with
$30\%$ contrast (Fig.~\ref{fig:fringes1}e)). To straighten our experimental fringes we first obtain, for each image row, the phase of the Fourier
component associated with the fringes (Fig.~\ref{fig:fringes1}b),f)). We then apply a linear phase fit across all rows of the Fourier transform,
before inverse Fourier transforming to obtain the images in Fig.~\ref{fig:fringes1}c),g). By averaging these corrected images over all rows, removing
the background and fitting sine curves to the experimental data we obtain the fringes and their contrast (Fig.~\ref{fig:fringes1}d),h)).

\begin{figure}[!th]
    \centering
        \includegraphics[width=.75\columnwidth]{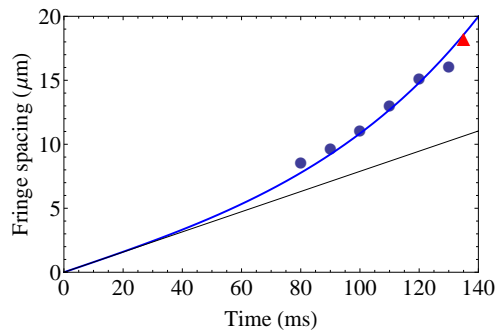}
    \caption{Fringe spacing as a function of levitation time. Ballistic expansion theory for $d=60\,\mu$m separated BECs (straight line) and
    experimental data (points) are shown as well as a blue $\sinh(\omega t)/\omega$ curve using $\omega=14\,$rad/s (a geometrical ring property),
    which has a $d=60\,\mu$m fit. The model aptly represents the fringe spacing in an inverted parabolic potential.
    The triangle data point is derived from the image in Fig.~\ref{fig:fringes1} \label{fig:FringeSpacing}}
\end{figure}

\begin{figure}[!t]
    \centering
        \includegraphics[width=\columnwidth]{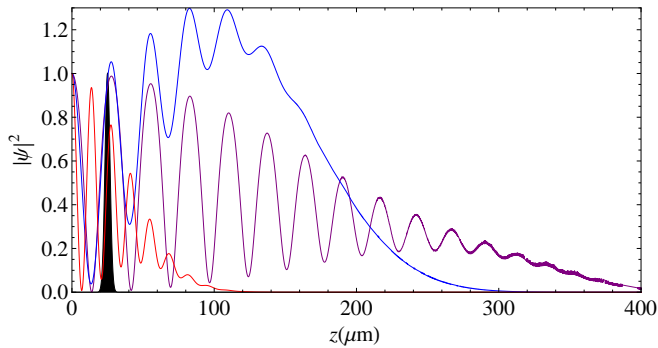}
    \caption{Relative theoretical probability distributions (fringes, with even symmetry about $z=0$) obtained when two Gaussian initial wavepackets (black) are released for $150\,$ms
    in a potential: $U_z=0$ (red), $U_z= -m\,\omega^2\, z^2/2$ (blue/purple curves). Interatomic repulsion is either absent (red/blue curves) or present (purple curve).
    Increasing the nonlinear term in the Schr\"{o}dinger equation affects the width of the final distribution, but not the fringe period\label{fig:FringeTheory}}
\end{figure}

Our BEC is levitated in an axial potential which is approximately an inverted parabola, $U_z= -m\,\omega^2\, z^2/2,$ due to the circular nature of
our toroidal quadrupole field. The magnitude of $\omega$ corresponds to that of a rigid pendulum, i.e. $\omega=g/r=14\,$rad/s where $g$ is the
acceleration due to gravity and $r=5\,$cm is the radius of our ring. By solving the one-dimensional, time-dependent, Schr\"{o}dinger equation (with
and without a nonlinear interatomic repulsion), one can show that the fringe spacing in the potential $U_z$ is modified from Eq.~\ref{deB} to
$\lambda'=\lambda \sinh(\omega t)/\omega.$ This interference fringe spacing dependence on the levitation potential is clearly observable in the
experimental fringe periodicity (Fig.~\ref{fig:FringeSpacing}). Theory also clearly shows shows that the fringe spacing is not altered by interatomic
repulsion (Fig.~\ref{fig:FringeTheory}). Interestingly the fringe separation has a similar dependence to that attributed to interatomic repulsion in
interferometry experiments on a chip, albeit at higher atomic density \cite{rfdress}.

The creation of spatial interference between split BECs with macroscopic separation offers a promising outlook for future atom interferometry based
measurements, e.g.\  our degenerate gas experiments in macroscopic ring geometries \cite{ring}. We intend to extend our proof-of-principle
experiments and perform interferometry with controlled phase by forming condensates with a weak link (due to lower dipole beam power) and raising the
barrier between condensates immediately before interferometric experiments. We will also carry out experiments with the plug's RF spectrum altered to
create BECs in multiple wells -- an `optical fork' for BECs, toward the limit of a 1D optical lattice with dynamic spacing. A weak carrier and small
separation between dipole beams will also allow the formation of a three-well BEC, ideal for `STIRAP' experiments \cite{parker} transferring a BEC
from the left quantum well (say) to the right quantum well, effectively bypassing the second quantum well.

This experiment was supported by the UK EPSRC and SUPA. PFG holds a RSE/Scottish Government Marie Curie Personal Research Fellowship.

\end{document}